# Electrochemical synthesis and crystal structure of the organic ion intercalated superconductor $(TMA)_{0.5}Fe_2Se_2$ with $T_c$ = 43 K

Bettina Rendenbach, Timotheus Hohl, Sascha Harm, Constantin Hoch, and Dirk Johrendt*

Department Chemie, Ludwig-Maximilians-Universität München, Butenandtstr. 5–13 (D), 81377 München, Germany, Email: johrendt@lmu.de

ABSTRACT Intercalation of organic cations in superconducting iron selenide can significantly increase the critical temperature ($T_c$). We present an electrochemical method using $\beta$-FeSe crystals ($T_c \approx 8$ K) floating on a mercury cathode to intercalate tetramethylammonium ions ($TMA^+$) quantitatively to obtain bulk samples of $(TMA)_{0.5}Fe_2Se_2$ with $T_c$ = 43 K. The layered crystal structure is closely related to the $ThCr_2Si_2$-type with disordered $TMA^+$ ions between the FeSe layers. Although the organic ions are not detectable by X-ray diffraction, packing requirements as well as first principle DFT calculations constrain the specified structure. Our synthetic route enables electrochemical intercalations of other organic cations with high yields to greatly optimize the superconducting properties, and to expand this class of high-$T_c$ materials.

In a superconductor electrons form pairs and electric currents flow dissipation-less below a critical temperature ($T_c$). Iron-based superconductors discovered in 2008[1, 2] represent the second class of high-temperature superconductors beyond the copper oxides, and attract tremendous interest equally in the physics and chemistry communities.[3] While superconducting wires based on copper oxides begin to capture the market in energy technologies,[4, 5] iron-based superconductors are still in an early stage of innovation.[6-8] Their main drawback is the critical temperature below the boiling point of liquid ammonia (77 K),[9] but in spite of immense efforts no bulk iron-based superconductor has reached the 77 K landmark so far. However, the finding of superconductivity up to 99 K[10] in thin $\beta$-FeSe films proved the potential for higher critical temperatures, once the right composition and structure is found. The common structural trait of all iron-based superconductors are layers of edge-sharing $FeX_{4/4}$ tetrahedra ($Fe^{2+}$, $X$ = As, Se), separated by interstitial atoms of various kinds. An impressive family of superconducting compounds[11] emerged by stacking of Fe$X$ layers with layers of alkaline,[12] alkaline-earth,[13] or rare-earth ions,[14] mixtures thereof,[15] or with thicker perovskite-like oxide layers.[16] A special case is the $\beta$-polymorph of iron selenide FeSe, which is a superconductor below 8 K without doping.[17] High pressure raises $T_c$ of bulk $\beta$-FeSe to 36.7 K at 8.9 GPa,[18] while one unit cell thin FeSe layers exhibit superconductivity up to 99 K when doped with electrons.[19,20] Likewise the critical temperature of bulk $\beta$-FeSe increases by electron transfer from cationic species in the van-der-Waals gap. Solid state reactions of $\beta$-FeSe with potassium yielded superconducting samples with $T_c$ around 30 K,[21] which are phase separated.[22]

However, intercalation reactions with $\beta$-FeSe as host compound can proceed at low temperatures via soft chemistry methods.[23-27] Examples are the intercalation of lithium ions with amine and amide species in liquid ammonia ($Li_x(NH_2)_y(NH_3)_{1-y}Fe_2Se_2$, $T_c$ = 43 K),[28] lithium hydroxide layers by hydrothermal methods ($[(Li_{1-x}Fe_x)OH]FeSe$, $T_c$ = 42 K)[29], or alkaline ions with amine molecules by solvothermal reactions ($Na_{0.39}(C_2N_2H_8)_{0.77}Fe_2Se_2$, $T_c$ = 46 K).[26]

Another promising approach is the electrochemical intercalation of alkali metal ions. Several studies reported electrochemically intercalated FeSe compounds with critical temperatures around 40±5 K.[30-34] Almost all of these materials suffer either from inhomogeneity, small superconducting volume fractions, or incomplete conversion of the host $\beta$-FeSe. Only recently, Shi et al. reported the intercalation of large cetyl-trimethylammonium ions ($CTA^+$)[35,36] and tetrabutylammonium ions ($TBA^+$)[37] in individual $\beta$-FeSe crystals with superconducting transitions up to 50 K. A drawback of this method is the tiny sample amount, consisting of one tiny crystal on the tip of an indium wire. Furthermore, the knowledge about the structures of the $CTA^+$ and $TBA^+$ intercalates is still incomplete and limited to the distance between the FeSe layers so far,[35,37] while the detailed structure, even of the FeSe layers therein, remains unknown. Here, we demonstrate the electrochemical intercalation of tetramethylammonium cations ($TMA^+$) into $\beta$-FeSe. A modified setup of the electrochemical cell yields single phase bulk samples of $(TMA)_{0.5}Fe_2Se_2$ with a superconducting transition at 43 K. We deduce a crystal structure closely related to the 122-type iron arsenide superconductors with $ThCr_2Si_2$-type structure.

Single crystals of the host $\beta$-FeSe were prepared by chemical vapor transport as described in the literature.[38-40] An optimized setup allows to grow about 1 gram $\beta$-FeSe crystals within one week. The quality of the host material was checked by powder X-ray diffraction of the crushed crystals and by magnetic susceptibility measurements. A portion of $\beta$-FeSe crystals was distributed on a drop of mercury in an amalgamated copper spoon serving as the cathode.[41] The crystals float on the surface of the mercury ensuring the electrical contact between the cathode and the electrolyte consisting of tetramethylammonium iodide (TMAI) dissolved in dry DMF. For details, we refer to the supporting information. During the electrolysis, the I$^-$ in the electrolyte is oxidized to I$_3^-$, while $\beta$-FeSe is reduced electrochemically with the charge compensated by the intercalation of TMA$^+$ ions. After the reaction is complete, the black air-sensitive crystals are easy to separate from the mercury drop. The yield scales with the size of the mercury surface and is about 50 to 200 mg per process cycle.

Figure 1 shows the powder X-ray pattern with the Rietveld fit. No impurity phases or residual $\beta$-FeSe are discernible within the limits of the method ($\approx$ 1 wt. %). Some intensities slightly deviate, which is caused by the preferred orientation of the plate-like crystallites.

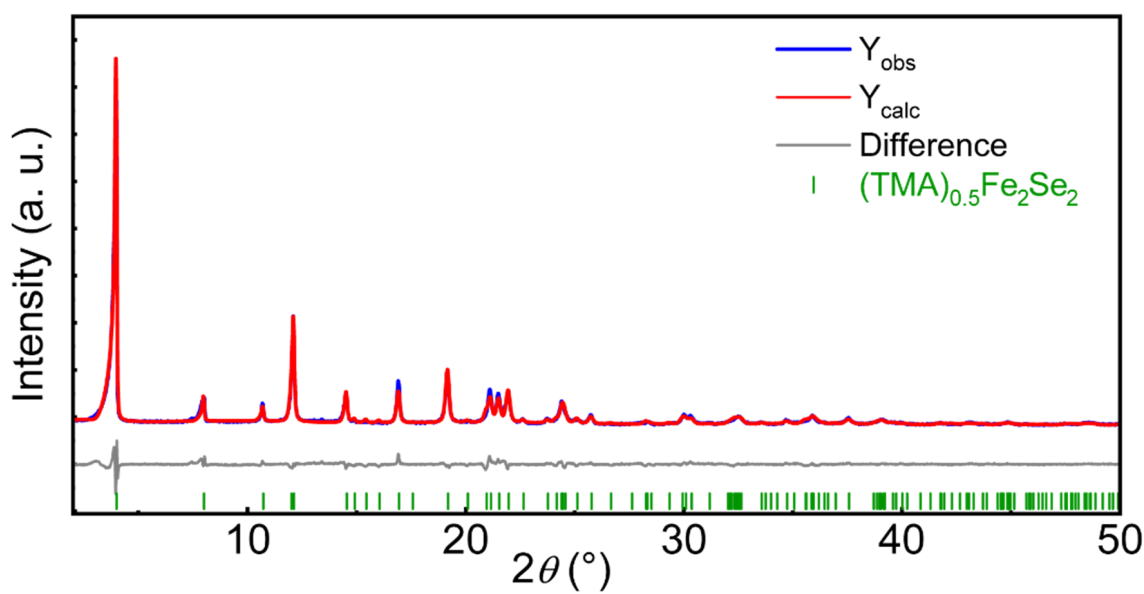

**Figure 1.** X-ray diffraction pattern of (TMA)$_{0.5}$Fe$_2$Se$_2$ (blue) with Rietveld fit (red) and difference curve (grey).

The crystal structure was solved from the powder X-ray diffraction data in the space group $I4/mmm$ with lattice parameters $a$ = 3.8585(2) Å, $c$ = 20.377(3) Å. Only the iron and selenium atoms contribute significantly to the diffraction pattern, because CHN atoms are weak scatterer and the orientations of the TMA$^+$ ions are very likely disordered. However, the nitrogen atom in the center of the TMA$^+$ ion is not affected by the disorder and was included in the refinement, though its contribution is expectedly weak. Relevant crystallographic data are compiled in Table 1.

**Table 1.** Crystallographic data of $(TMA)_{0.5}Fe_2Se_2$

| | |
|---|---|
| Space group | $I4/mmm$ (139) |
| Lattice parameters (Å) | $a$ = 3.8585(2) |
| | $c$ = 20.377(3) |
| Volume (Å$^3$) | $V$ = 303.4(1) |
| Z | 2 |
| Density (g·cm$^{-3}$) | 3.03(1) |
| $R_{exp}/R_{Bragg}$ | 0.815/1.269 |
| $R_p/R_{wp}$ | 1.940/2.789 |
| GooF | 3.423 |

Atomic coordinates and displacement parameters

| Atom | Wyckoff | x | y | z | SOF | $B_{iso}$ |
|---|---|---|---|---|---|---|
| Fe | 4d | 0 | ½ | ¼ | 1 | 2.0(1) |
| Se | 4e | 0 | 0 | 0.3160(4) | 1 | 1.2(1) |
| N | 2a | 0 | 0 | 0 | 0.5 | 3.9(1) |

Bond distances (Å) and angles (°)

Fe – 4 Se   2.352(3)         φ Se-Fe-Se   110.2(3)

This atom configuration in the space group $I4/mmm$ corresponds to the ThCr$_2$Si$_2$-type structure, known as the "122-type" structure of the iron arsenide superconductors.[13] Even though the cavities in the structure around the N-atom sites at (0,0,0) and (½,½,½) appear large (Figure 2a), they are not large enough to be fully occupied by TMA$^+$ ions. The encasing sphere of a tetrahedrally shaped TMA$^+$ ion has a diameter of 5.5–5.6 Å[42, 43] and is therefore incompatible with the lattice parameter $a$ = 3.8585(2) Å.

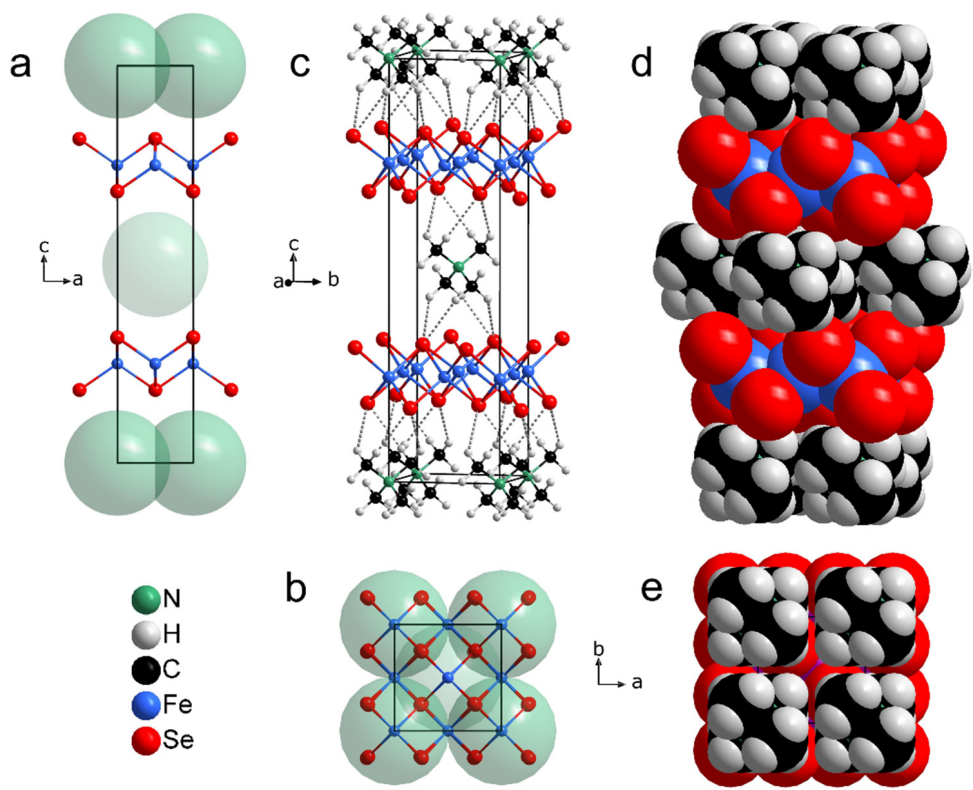

**Figure 2**. Crystal structure of (TMA)$_{0.5}$Fe$_2$Se$_2$ **a**: Structure determined from powder X-ray diffraction in space group $I4/mmm$. Large green spheres indicate the space required by a TMA$^+$ ion. **b**: Doubled unit cell with $a'=\sqrt{2}a$, $b'=\sqrt{2}b$ with perfectly fitting TMA spheres. **c**: Complete structure of (TMA)$_{0.5}$Fe$_2$Se$_2$ in space group $I\bar{4}2m$ with hydrogen bridges shown as dashed bonds. **d**: Space filling model **e**: View perpendicular to the TMA$^+$ layers.

Indeed the diagonal of the unit cell $\sqrt{2}a$ = 5.457 Å has the suitable size to acccomodate neighboring TMA$^+$ ions (Figure 2b), thus we assume that the positions (0,0,0) and (½,½,½) are statistically half occupied, resulting in the formula (TMA)$_{0.5}$Fe$_2$Se$_2$. An ordered model of the structure in space group $I\bar{4}2m$ is shown in Figure 2c as ball-and-stick representation, and in Figure 2d with the van-der-Waals radii of the atoms. Figure 2e shows how the TMA$^+$ ions almost perfectly fit in the $\sqrt{2}a \times \sqrt{2}a$ supercell. The space filling of this structure as calculated by PLATON[44] is as high as 77 %, similar to a typical compound with ThCr$_2$Si$_2$-type structure like BaFe$_2$As$_2$ which has a space filling of 82 %. Each two of the three hydrogen atoms at the -CH$_3$ groups form C-H···Se hydrogen bridges with a H···Se distance of 2.72 Å (Figure 2c), similar to the N-D···Se distance of 2.76 Å measured by Burrard-Lucas et al. in Li$_x$(ND$_2$)$_y$(ND$_3$)$_{1-y}$Fe$_2$Se$_2$ using neutron diffraction.[28] We do not expect more accurate structural data from neutron diffraction because of the orientational disorder of the TMA$^+$ molecules. In comparable compounds like Na$_{0.39}$(C$_2$N$_2$H$_8$)$_{0.77}$Fe$_2$Se$_2$, the molecules could not be localized by neutron diffraction either.[26]

We have checked the validity of this model by first principle DFT calculations using the VASP code.[45-47] DFT reproduces experimental structures within a certain accuracy that depends on the functional used. We have chosen the SCAN[48] functional, which reproduces the experimental lattice parameters of our compound within 0.1 %. The Fe-Se bond length and Se-Fe-Se bond angle deviate by only +2.2 % and +0.4 % from the experimental values, respectively. Table 2 shows the calculated structure data in the space group $I\bar{4}2m$ with experimental values in square brackets. Table S1 compares the experimental parameters with calculated ones using different exchange-correlation functionals. The excellent agreement with the experimental values clearly supports our structure model.

**Table 2.** Calculated structure parameters of $(TMA)_{0.5}Fe_2Se_2$ with experimetal values in square brackets.

| Crystal system | Tetragonal | | |
|---|---|---|---|
| Space group | $I\bar{4}2m$ (No. 121) | | |
| Lattice parameters (Å) | $a$ = 5.454 [5.457] | | |
| | $c$ = 20.383 [20.377] | | |
| Volume (Å$^3$) | 606.3 [606.7] | | |
| Atom positions | | | |
| Atom | Wyckoff | $x$ | $y$ | $z$ |
| Se | 8$i$ | 0.7570 [¾] | $x$ | 0.6797 [0.6840] |
| Fe1 | 4$e$ | 0 | 0 | 0.7500 [ ¾ ] |
| Fe2 | 4$d$ | 0 | ½ | ¼ [ ¼ ] |
| N | 2$a$ | 0 [0] | 0 [0] | 0 [0] |
| C | 8$i$ | 0.1595 | $x$ | 0.9583 |
| H1 | 8$i$ | 0.2706 | $x$ | 0.9912 |
| H2 | 16$j$ | 0.0435 | 0.7270 | 0.0729 |

To get an idea of the thermodynamic stability, we have calculated the phonon dispersions and phonon density-of-states. A crystal is stable if its potential energy increases against any combinations of atomic displacements, which means that all phonons have real (positive) frequencies.[49] Calculations using the space group $I\bar{4}2m$ reveal only minor imaginary modes (Figure S1a). This reflects the fact that our structure model is not perfect and neglects the disorder of the organic cations, which is not treatable by DFT methods. The TMA$^+$ ions are possibly ordered over half the sites in the layers due to space requirements, but order is lost along the $c$ axis. Interestingly, no imaginary modes occur after structure optimizations without symmetry constraints (Figure S1b). Even though this triclinic structure is chemically reasonable, its unit cell is incompatible with the X-ray diffraction pattern.

Chemical C-H-N analysis, EDS, and FT-IR spectroscopy confirm the composition (TMA)$_{0.5}$Fe$_2$Se$_2$ within the errors of these methods, respectively. For details, we refer to the supplementary information. The intercalation is topotactic and reversible. High-temperature powder X-ray diffraction reveals the complete recovery of tetragonal $\beta$-FeSe after heating to 200 °C, and the transformation to the hexagonal polymorph at 550 °C (Figure 3).

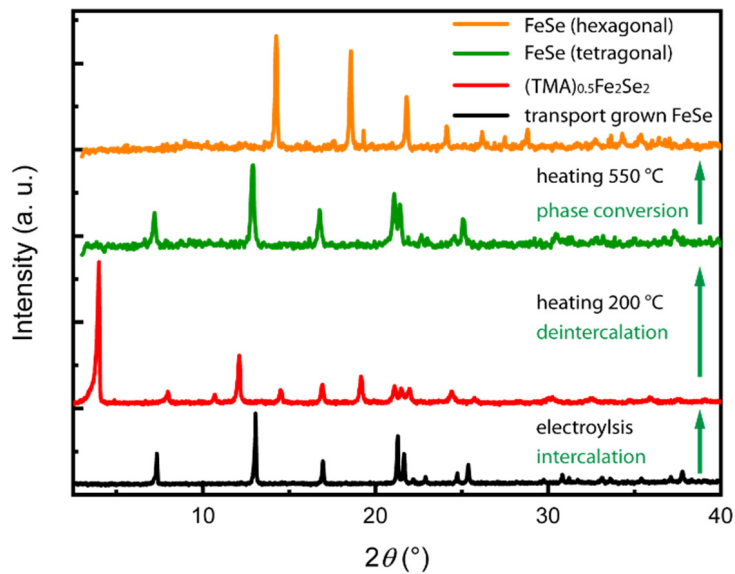

**Figure 3.** Powder X-ray diffraction patterns (Mo-K$_{\alpha 1}$ radiation) of the host $\beta$-FeSe (black), (TMA)$_{0.5}$Fe$_2$Se$_2$ after electrochemical intercalation (red), recovered $\beta$-FeSe after deintercalation (green) and after conversion to hexagonal FeSe (orange).

The magnetic susceptibility of (TMA)$_{0.5}$Fe$_2$Se$_2$ shows a bulk superconducting transition at 43 K (Figure 4). Field-cooled and zero-field cooled curves slightly split above $T_c$ due to traces of ferromagnetic impurities not detectable by X-ray diffraction.

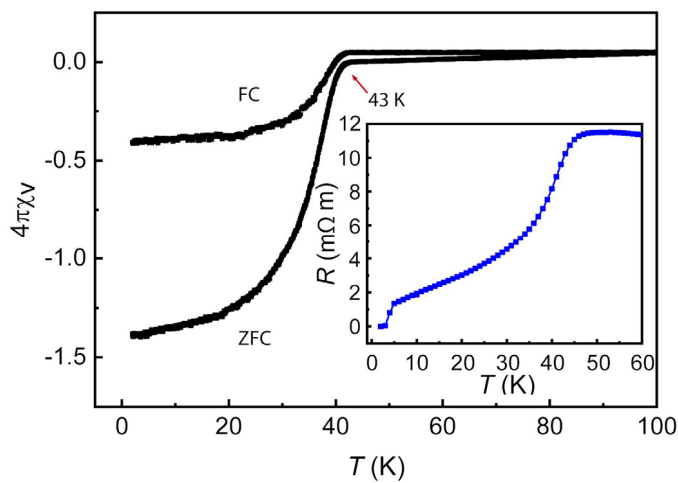

**Figure 4.** Susceptibility of (TMA)$_{0.5}$Fe$_2$Se$_2$ at 15 Oe. Insert: DC resistivity of a cold pressed pellet.

The large shielding fraction above 100 % at low temperatures comes from the uncorrected demagnetization of the plate-like crystallites oriented perpendicular to the magnetic field. No further drop of the susceptibility near 8 K is visible, which confirms that the intercalation is complete and no residual host β-FeSe remains. Measuring the electrical resistivity turned out difficult due to degradation of the sample during pressing, and furthermore the deintercalation temperature around 200 °C allowed no sintering of the pellet. The result is shown in the insert of Figure 4, where the onset of the superconductivity is near 45 K followed by a broad transition until an additional drop to zero resistivity occurs at 6 K. The latter is caused by ~8 wt.-% deintercalated FeSe. Isothermal magnetization measurements (Figure 5) show the "butterfly" pattern typical for a hard type-II superconductor. The ripples in the curve only occur at increasing field, which indicates that they are flux jumps.

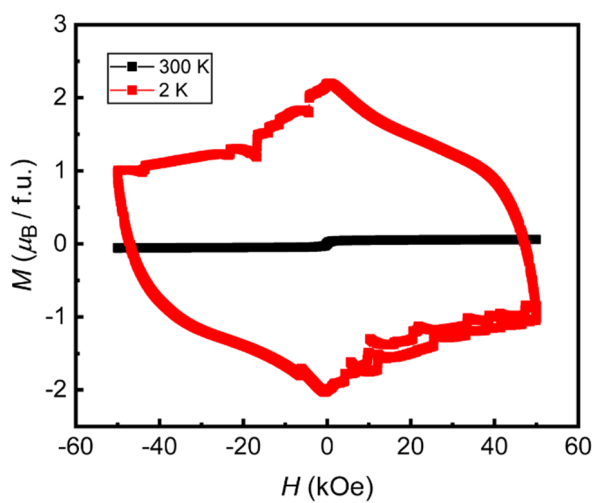

**Figure 5.** Magnetization isotherms of (TMA)$_{0.5}$Fe$_2$Se$_2$ at 2 K and at 300 K.

In conclusion, we demonstrate that the electrochemical intercalation of tetramethylammonium ions (TMA$^+$) into the van-der-Waals gap of $\beta$-FeSe is feasible with high yields. Powder X-ray diffraction combined with DFT calculations reveal a reliable ordered model of the crystal structure. (TMA)$_{0.5}$Fe$_2$Se$_2$ forms a variant of the ThCr$_2$Si$_2$-type structure, also known as the "122-type" in the family of iron arsenide superconductors. The TMA$^+$ ions are closely packed between the FeSe layers but disordered over two equivalent positions and in different orientations. Magnetization and electrical resistivity measurements show bulk superconducting transitions at 43 K and identify (TMA)$_{0.5}$Fe$_2$Se$_2$ as type-II superconductor. Our results provide the first insights into the crystal structure of a superconducting FeSe-alkylammonium intercalate and pave the way to further exploit the electrochemical route towards related compounds with potentially higher critical temperatures.

## ASSOCIATED CONTENT

**Supporting Information**

Details about the synthesis, structure determination by powder X-ray powder diffraction, DFT calculations, chemical analysis, magnetic measurements, and IR spectroscopy.

## AUTHOR INFORMATION

**Corresponding Author**

Prof. Dr. Dirk Johrendt, Department Chemie, Ludwig-Maximilians-Universität München, Butenandtstr. 5–13 (Haus D), 81377 München, Germany

**Author Contributions**

The manuscript was written through contributions of all authors. All authors have given approval to the final version of the manuscript.

**Funding Sources**

We gratefully acknowledge financial support by the Deutsche Forschungsgemeinschaft (DFG), grant No. JO257/7-2..

## ACKNOWLEDGMENT

The authors thank the German Research Foundation (DFG- Deutsche Forschungsgemeinschaft) for funding this work (grant JO257/7-2). We thank Valentin Weippert and Tobias Rackl for the magnetic and resistivity measurements, Alan Virmani for the FT-IR measurements, and Ralph Stoffel from Richard Dronskowski's group at RWTH Aachen for valuable help with the VASP calculations. The authors gratefully acknowledge the computational resources provided by the Leibniz Supercomputing Centre (www.lrz.de).

## REFERENCES

1. Kamihara, Y.; Watanabe, T.; Hirano, M.; Hosono, H., Iron-Based Layered Superconductor La[O1-xFx]FeAs (x = 0.05-0.12) with Tc = 26 K. *J. Am. Chem. Soc.* **2008**, *130*, (11), 3296-3297.

2. Rotter, M.; Pangerl, M.; Tegel, M.; Johrendt, D., Superconductivity and Crystal Structures of $(Ba_{1-x}K_x)Fe_2As_2$ (x = 0-1). *Angew. Chem. Int. Ed.* **2008**, *47*, (41), 7949-7952.

3. Si, Q.; Yu, R.; Abrahams, E., High-temperature superconductivity in iron pnictides and chalcogenides. *Nat. Rev. Mater.* **2016**, *1*, (4), 16017.

4. Bergen, A.; Andersen, R.; Bauer, M.; Boy, H.; Brake, M. t.; Brutsaert, P.; Bührer, C.; Dhallé, M.; Hansen, J.; ten Kate, H.; Kellers, J.; Krause, J.; Krooshoop, E.; Kruse, C.; Kylling, H.; Pilas, M.; Pütz, H.; Rebsdorf, A.; Reckhard, M.; Seitz, E.; Springer, H.; Song, X.; Tzabar, N.; Wessel, S.; Wiezoreck, J.; Winkler, T.; Yagotyntsev, K., Design and in-field testing of the world's first ReBCO rotor for a 3.6 MW wind generator. *Supercond. Sci. Technol.* **2019**, *32*, (12), 125006.

5. *High Temperature Superconductors (HTS) for Energy Applications*, Melhem, Z. E., Ed.; Woodhead Publishing Ltd.; Sawston, Cambridge, UK, **2012**.

6. Yao, C.; Ma, Y., Recent breakthrough development in iron-based superconducting wires for practical applications. *Supercond. Sci. Technol.* **2019**, *32*, (2), 023002.

7. Hosono, H.; Tanabe, K.; Takayama-Muromachi, E.; Kageyama, H.; Yamanaka, S.; Kumakura, H.; Nohara, M.; Hiramatsu, H.; Fujitsu, S., Exploration of new superconductors and functional materials, and fabrication of superconducting tapes and wires of iron pnictides. *Sci. Technol. Adv. Mater.* **2015**, *16*, (3), 033503.

8. Hosono, H.; Yamamoto, A.; Hiramatsu, H.; Ma, Y., Recent advances in iron-based superconductors toward applications. *Mater. Today* **2018**, *31*, (3), 278-302.

9. Zestrea, V.; Kodash, V. Y.; Felea, V.; Petrenco, P.; Quach, D. V.; Groza, J. R.; Tsurkan, V., Structural and magnetic properties of $FeCr_2S_4$ spinel prepared by field-activated sintering and conventional solid-state synthesis. *J. Mater. Sci.* **2008**, *43*, (2), 660-664.

10. Ge, J.-F.; Liu, Z.-L.; Liu, C.; Gao, C.-L.; Qian, D.; Xue, Q.-K.; Liu, Y.; Jia, J.-F., Superconductivity above 100 K in single-layer FeSe films on doped $SrTiO_3$. *Nat. Mater.* **2015**, *14*, (3), 285–289.

11. Stewart, G. R., Superconductivity in iron compounds. *Rev. Mod. Phys.* **2011**, *83*, (4), 1589-1652.

12. Tapp, J. H.; Tang, Z. J.; Lv, B.; Sasmal, K.; Lorenz, B.; Chu, P. C. W.; Guloy, A. M., LiFeAs: An intrinsic FeAs-based superconductor with Tc = 18 K. *Phys. Rev. B* **2008**, *78*, (6), 060505.

13. Rotter, M.; Tegel, M.; Johrendt, D., Superconductivity at 38 K in the iron arsenide $(Ba_{1-x}K_x)Fe_2As_2$. *Phys. Rev. Lett.* **2008**, *101*, (10), 107006.

14. Ren, Z. A.; Che, G. C.; Dong, X. L.; Yang, J.; Lu, W.; Yi, W.; Shen, X. L.; Li, Z. C.; Sun, L. L.; Zhou, F.; Zhao, Z. X., Superconductivity and phase diagram in iron-based arsenic-oxides ReFeAsO(1-δ) (Re = rare-earth metal) without fluorine doping. *Europhys. Lett.* **2008**, *83*, (1), 4.

15. Iyo, A.; Kawashima, K.; Kinjo, T.; Nishio, T.; Ishida, S.; Fujihisa, H.; Gotoh, Y.; Kihou, K.; Eisaki, H.; Yoshida, Y., New-Structure-Type Fe-Based Superconductors: $CaAFe_4As_4$ (A = K, Rb, Cs) and $SrAFe_4As_4$ (A = Rb, Cs). *J. Am. Chem. Soc.* **2016**, *138*, (10), 3410–3415.

16. Zhu, X.; Han, F.; Mu, G.; Cheng, P.; Shen, B.; Zeng, B.; Wen, H.-H., Transition of stoichiometric $Sr_2VO_3FeAs$ to a superconducting state at 37.2 K. *Phys. Rev. B* **2009**, *79*, (22), 220512-4.

17. Hsu, F.-C.; Luo, J.-Y.; Yeh, K.-W.; Chen, T.-K.; Huang, T.-W.; Wu, P. M.; Lee, Y.-C.; Huang, Y.-L.; Chu, Y.-Y.; Yan, D.-C.; Wu, M.-K., Superconductivity in the PbO-type structure α-FeSe. *Proc. Natl. Acad. Sci. U.S.A.* **2008**, *105*, (38), 14262–14264.

18. Medvedev, S.; McQueen, T. M.; Troyan, I. A.; Palasyuk, T.; Eremets, M. I.; Cava, R. J.; Naghavi, S.; Casper, F.; Ksenofontov, V.; Wortmann, G.; Felser, C., Electronic and magnetic phase diagram of beta-$Fe_{1.01}$Se with superconductivity at 36.7 K under pressure. *Nat. Mater.* **2009**, *8*, (8), 630-633.

19. Zhou, Y. J.; Millis, A. J., Charge transfer and electron-phonon coupling in monolayer FeSe on Nb-doped $SrTiO3$. *Phys. Rev. B* **2016**, *93*, (22), 8.

20. Miyata, Y.; Nakayama, K.; Sugawara, K.; Sato, T.; Takahashi, T., High-temperature superconductivity in potassium-coated multilayer FeSe thin films. *Nat. Mater.* **2015**, *14*, (8), 775–779.

21. Guo, J. G.; Jin, S. F.; Wang, G.; Wang, S. C.; Zhu, K. X.; Zhou, T. T.; He, M.; Chen, X. L., Superconductivity in the iron selenide $K_xFe_2Se_2$ (0 <= x <= 1.0). *Phys. Rev. B* **2010**, *82*, (18), 180520.

22. Shoemaker, D. P.; Chung, D. Y.; Claus, H.; Francisco, M. C.; Avci, S.; Llobet, A.; Kanatzidis, M. G., Phase relations in $K_xFe_{2-y}Se_2$ and the structure of superconducting $K_xFe_2Se_2$ via high-resolution synchrotron diffraction. *Phys. Rev. B* **2012**, *86*, (18), 184511.

23. Krzton-Maziopa, A.; Pesko, E.; Puzniak, R., Superconducting selenides intercalated with organic molecules: synthesis, crystal structure, electric and magnetic properties, superconducting properties, and phase separation in iron based-chalcogenides and hybrid organic-inorganic superconductors. *J. Phys. Condens. Matter* **2018**, *30*, (24), 243001.

24. Dong, X.; Zhou, H.; Yang, H.; Yuan, J.; Jin, K.; Zhou, F.; Yuan, D.; Wei, L.; Li, J.; Wang, X.; Zhang, G.; Zhao, Z., Phase Diagram of $(Li_{1-x}Fe_x)OHFeSe$: A Bridge between Iron Selenide and Arsenide Superconductors. *J. Am. Chem. Soc.* **2015**, *137*, (1), 66-69.

25. Sun, R.; Jin, S.; Gu, L.; Zhang, Q.; Huang, Q.; Ying, T.; Peng, Y.; Deng, J.; Yin, Z.; Chen, X., Intercalating Anions between Terminated Anion Layers: Unusual Ionic S–Se Bonds and Hole-Doping Induced Superconductivity in $S_{0.24}(NH_3)_{0.26}Fe_2Se_2$. *J. Am. Chem. Soc.* **2019**, *141*, (35), 13849–13857.

26. Jin, S.; Fan, X.; Wu, X.; Sun, R.; Wu, H.; Huang, Q.; Shi, C.; Xi, X.; Li, Z.; Chen, X., High-Tc superconducting phases in organic molecular intercalated iron selenides: synthesis and crystal structures. *Chem. Commun.* **2017**, *53*, (70), 9729-9732.

27. Xu, H.-S.; Wang, X.-X.; Tang, L.-L.; Yang, K.-P.; Yang, D.; Long, Y.-Z.; Tang, K.-B., New Synthetic Route to Synthesize Li and 1,2-Diaminopropane-Intercalated Iron-Based Superconductor with Tc = 37 K. *ChemistrySelect* **2018**, *3*, (27), 7757–7762.

28. Burrard-Lucas, M.; Free, D. G.; Sedlmaier, S. J.; Wright, J. D.; Cassidy, S. J.; Hara, Y.; Corkett, A. J.; Lancaster, T.; Baker, P. J.; Blundell, S. J.; Clarke, S. J., Enhancement of the superconducting transition temperature of FeSe by intercalation of a molecular spacer layer. *Nat. Mater.* **2013**, *12*, 15-19.

29. Pachmayr, U.; Nitsche, F.; Luetkens, H.; Kamusella, S.; Brückner, F.; Sarkar, R.; Klauss, H.-H.; Johrendt, D., Coexistence of 3d-Ferromagnetism and Superconductivity in $[(Li_{1-x}Fe_x)OH](Fe_{1-y}Li_y)Se$. *Angew. Chem. Int. Ed.* **2015**, *54*, (1), 293-297.

30. Abe, H.; Noji, T.; Kato, M.; Koike, Y., Electrochemical Li-intercalation into the Fe-based superconductor $FeSe_{1-x}Te_x$. *Physica C* **2010**, *470*, S487–S488.

31. Kajita, T.; Kawamata, T.; Noji, T.; Hatakeda, T.; Kato, M.; Koike, Y.; Itoh, T., Electrochemical Na-intercalation-induced high-temperature superconductivity in FeSe. *Physica C* **2015**, *519*, 104-107.

32. Shen, S.-J.; Ying, T.-P.; Wang, G.; Jin, S.-F.; Zhang, H.; Lin, Z.-P.; Chen, X.-L., Electrochemical synthesis of alkali-intercalated iron selenide superconductors. *Chin. Phys. B* **2015**, *24*, (11), 117406.

33. Alekseeva, A. M.; Drozhzhin, O. A.; Dosaev, K. A.; Antipov, E. V.; Zakharov, K. V.; Volkova, O. S.; Chareev, D. A.; Vasiliev, A. N.; Koz, C.; Schwarz, U.; Rosner, H.; Grin, Y., New superconductor $Li_xFe_{1+\delta}Se$ (x ≤ 0.07, Tc up to 44 K) by an electrochemical route. *Sci. Rep.-UK* **2016**, *6*, (1), 25624.

34. Cui, Y.; Hu, Z.; Zhang, J.-S.; Ma, W.-L.; Ma, M.-W.; Ma, Z.; Wang, C.; Yan, J.-Q.; Sun, J.-P.; Cheng, J.-G.; Jia, S.; Li, Y.; Wen, J.-S.; Lei, H.-C.; Yu, P.; Ji, W.; Yu, W.-Q., Ionic-Liquid-Gating Induced Protonation and Superconductivity in FeSe, FeSe0.93S0.07, ZrNCl, 1T-TaS2 and Bi2Se3 *Chin. Phys. Lett.* **2019**, *36*, (7), 077401.

35. Shi, M. Z.; Wang, N. Z.; Lei, B.; Shang, C.; Meng, F. B.; Ma, L. K.; Zhang, F. X.; Kuang, D. Z.; Chen, X. H., Organic-ion-intercalated FeSe-based superconductors. *Phys. Rev. Mater.* **2018**, *2*, (7), 074801.

36. Sun, J. P.; Shi, M. Z.; Lei, B.; Xu, S. X.; Uwatoko, Y.; Chen, X. H.; Cheng, J. G., Pressure-induced second high-Tc superconducting phase in the organic-ion-intercalated $(CTA)_{0.3}FeSe$ single crystal. *Europhys. Lett.* **2020**, *130*, (6), 67004.

37. Shi, M. Z.; Wang, N. Z.; Lei, B.; Ying, J. J.; Zhu, C. S.; Sun, Z. L.; Cui, J. H.; Meng, F. B.; Shang, C.; Ma, L. K.; Chen, X. H., FeSe-based superconductors with a superconducting transition temperature of 50 K. *New J. Phys.* **2018**, *20*, (12), 123007.

38. Böhmer, A. E.; Hardy, F.; Eilers, F.; Ernst, D.; Adelmann, P.; Schweiss, P.; Wolf, T.; Meingast, C., Lack of coupling between superconductivity and orthorhombic distortion in stoichiometric single-crystalline FeSe. *Phys. Rev. B* **2013**, *87*, (18), 180505.

39. Böhmer, A. E.; Taufour, V.; Straszheim, W. E.; Wolf, T.; Canfield, P. C., Variation of transition temperatures and residual resistivity ratio in vapor-grown FeSe. *Phys. Rev. B* 2016, *94*, (2), 024526.

40. Chareev, D.; Osadchii, E.; Kuzmicheva, T.; Lin, J.-Y.; Kuzmichev, S.; Volkova, O.; Vasiliev, A., Single crystal growth and characterization of tetragonal $FeSe_{1-x}$ superconductors. *CrystEngComm* **2013**, *15*, (10), 1989–1993.

41. Tambornino, F.; Sappl, J.; Pultar, F.; Cong, T. M.; Hübner, S.; Giftthaler, T.; Hoch, C., Electrocrystallization: A Synthetic Method for Intermetallic Phases with Polar Metal–Metal Bonding. *Inorg. Chem.* **2016**, *55*, (21), 11551–11559.

42. Gao, Q.; Giraldo, O.; Tong, W.; Suib, S. L., Preparation of Nanometer-Sized Manganese Oxides by Intercalation of Organic Ammonium Ions in Synthetic Birnessite OL-1. *Chem. Mater.* **2001**, *13*, (3), 778–786.

43. Liu, Z.-h.; Wang, Z.-M.; Yang, X.; Ooi, K., Intercalation of Organic Ammonium Ions into Layered Graphite Oxide. *Langmuir* **2002**, *18*, (12), 4926–4932.

44. Spek, A., Single-crystal structure validation with the program PLATON. *J. Appl. Crystallogr.* **2003**, *36*, (1), 7-13.

45. Kresse, G.; Furthmüller, J., Efficiency of ab-initio total energy calculations for metals and semiconductors using a plane-wave basis set. *Comput. Mat. Sci.* **1996**, *6*, (1), 15-50.

46. Kresse, G.; Furthmüller, J., Efficient iterative schemes for ab initio total-energy calculations using a plane-wave basis set. *Phys. Rev. B* **1996**, *54*, (16), 11169-11186.

47. Kresse, G.; Joubert, D., From ultrasoft pseudopotentials to the projector augmented-wave method. *Phys. Rev. B* **1999**, *59*, (3), 1758-1775.

48. Sun, J.; Ruzsinszky, A.; Perdew, J. P., Strongly Constrained and Appropriately Normed Semilocal Density Functional. *Phys. Rev. Lett.* **2015**, 115, (3), 036402.

49. Togo, A.; Tanaka, I., First principles phonon calculations in materials science. S*cripta Mater.* **2015**, *108*, 1-5.

# Supporting Information

## 1. Synthesis

(TMA)$_{0.5}$Fe$_2$Se$_2$ (TMA = tetramethylammonium, NC$_4$H$_{12}^+$) was syntheszied by electrochemical intercalation of tetramethylammonium iodide into iron selenide crystals. The host compound $\beta$-FeSe was synthesized with chemical vapour-transport.[1] 562.4 mg Se powder (Chempur, 99.9 %) and 437.8 mg Fe powder (CHEMPUR, 99.9 %) in a molar ratio 1 : 1.1 were ground together with AlCl$_3$/KCl (ALFA AESAR, 99.985 % / GRÜSSING, 99.5 %, dried) (7.75 g : 2.25 g). The mixture was sealed under vacuum in a glass ampoule (diameter 5 cm, length 4 cm) and placed in a vertical two-zone furnace and heated to 390 °C at the bottom and 290 °C at the top. This temperature gradient was held for 5–10 days.[2] After cooling the ampoules were opened and the crystals collected from the inner top of the ampoules.

The electrochemical cell consisted of a tungsten anode and an amalgamated copper spoon connected to a platinum wire as cathode. A drop of mercury was added to the spoon.[2] On top of this drop the polycrystalline powder of $\beta$-FeSe was distributed. The apparatus was held under inert conditions using purified argon. The electrolyte consisted of tetramethylammonium iodide (TMAI, SIGMA-ALDRICH, 99 %, 0.1 molar) dissolved in 100 mL dried and destilled DMF. A voltage of 3 V was applied for 3–4 days. After the reaction the product was washed with dry DMF and dried under vaccuum.

## 2. Powder X-ray diffraction

Glass capillaries (0.3 mm in diameter, Hilgenberg GmbH) were filled with the samples and sealed. A Stoe Stadi-P diffracometer (Mo$_{K\alpha1}$, Ge(111)-monochromator, Mythen 1k detector) was used to measure the patterns which were analysed and fitted using the Topas package.[3,4] After indexing the data with the SVG-algorithm, the space group *I*4/*mmm* was chosen.[5] Intensities were gathered using the Pawley method, and the structure was solved by charge-flipping.[3,6,7] The trial structures were used in subsequent Rietveld refinements and visualized by the program Diamond.[8] Measurements at high temperatures performed on samples in silicia capillaries (diameter 0.5 mm, Hilgenberg GmbH, sealed with grease) on a Stoe Stadi-P diffractometer (Mo$_{K\alpha}$, Ge(111)-monochromator, IP-PSD detector) equipped with a graphite furnace. Data were visualized with WinXPOW.[9]

### 3. DFT calculations

First-principles electronic structure calculations were performed using the Vienna ab initio simulation package (VASP 5.4.4)[10, 11] based on density functional theory (DFT) and plane wave basis sets. Projector-augmented waves (PAW)[12] were used and contributions of correlation and exchange were treated using the strongly constrained and appropriately normed semi-local density functional (SCAN).[13] The $k$-space was sampled with the Monkhorst-Pack[14] scheme using an $11 \times 11 \times 11$ grid based on the primitive unit cell. The AFLOW[15] utilities were used to transform between primitive and conventional unit cells, and FINDSYM[16] to determine the space group symmetry. Convergence criteria were $10^{-8}$ eV for the total energy and $10^{-4}$ eV/Å for the structural relaxations regarding ion positions, respectively, using a plane wave cut-off energy of 600 eV. The parameters of the fully relaxed structure of $(TMA)_{0.5}Fe_2Se_2$ in the space group $I\bar{4}2m$ are compiled in Table S1.

**Table S1.** Structure parameters of $(TMA)_{0.5}Fe_2Se_2$ (SG $I\bar{4}2m$) calculated with different functionals

|  | $a$ (Å) | Δ % | $c$ (Å) | Δ % | $V$ (Å$^3$) | Δ % | $z_{SE}$ | Δ% | Fe-Se | Δ % | Se-Fe-Se | Δ % |
|---|---|---|---|---|---|---|---|---|---|---|---|---|
| Exptl. | 5.457 |  | 20.377 |  | 606.75 |  | 0.3160 |  | 2.352 |  | 110.24 |  |
| SCAN | 5.454 | -0.05 | 20.383 | 0.03 | 606.32 | -0.07 | 0.3203 | 1.36 | 2.403 | 2.18 | 110.62 | 0.35 |
| PBE | 5.393 | -1.16 | 20.7151 | 1.66 | 602.62 | -0.67 | 0.3175 | 0.47 | 2.365 | 0.55 | 105.04 | -4.71 |
| PBEsol | 5.260 | -3.60 | 20.0346 | -1.68 | 554.36 | -8.63 | 0.3180 | 0.62 | 2.306 | -1.97 | 110.55 | 0.28 |
| LDA | 5.189 | -4.91 | 19.3535 | -5.02 | 521.07 | -14.12 | 0.3178 | 0.58 | 2.256 | -4.06 | 111.88 | 1.48 |

The phonon dispersions and phonon DOS shown in Figure S1 were calculated from forces acting on displaced atoms in 2×2×2 supercells using PHONOPY[17] and plotted with the SUMO tools.[18] The structural parameters were previously optimized until all forces were smaller than $10^{-5}$ eV/Å and energy changes are below $10^{-9}$ eV. This fully relaxed structure was used to calculate the infrared absorptions using density functional perturbation theory (DFPT, Figure S3).

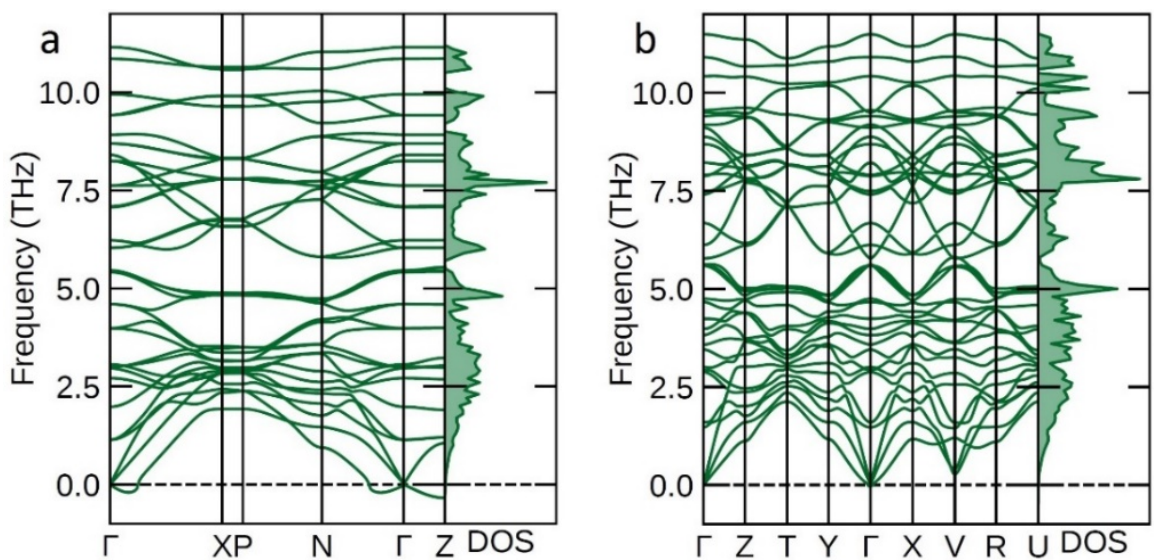

**Figure S1.** Phonon dispersion and DOS of $(TMA)_{0.5}Fe_2Se_2$. (a) Structure in space group $I\bar{4}2m$ (b) Triclinic structure in $P1$ after optimization without symmetry constraints.

## 4. Chemical analysis

CHN elemental analysis and energy-dispersive spectroscopy measurements (EDS) confirm the chemical composition $(TMA)_{0.5}Fe_2Se_2$ (see Tables S2 and S3). Figure S2 shows the morphology of a crystallite after intercalation. EDS measurments were performed on a Carl Zeiss Evo-Ma10 microscope with a Bruker Nano EDX detector (X-Flash detector 410 M). The controlling software is SmartSem for the detectors (SE and BSE)[19] and for the collections and evaluation of the spectra the program QUANTAX 200 was used.[20] Any elements from the sample holder and the adhesive carbon pads were discounted.

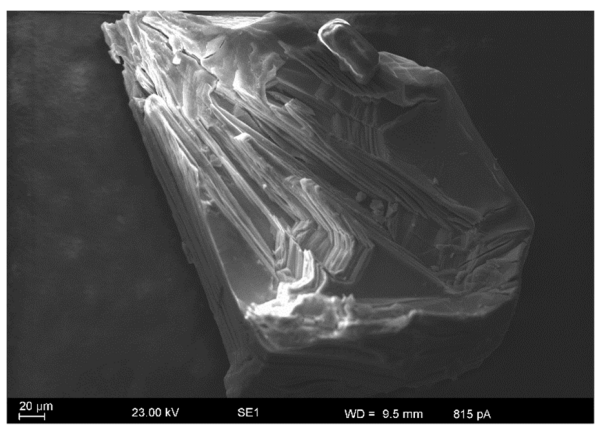

**Figure S2.** SEM image of $(TMA)_{0.5}Fe_2Se_2$.

**Table S2.** C:N:H ratios from elemental analysis normalized to N = 1.

|  | C | H | N |
|---|---|---|---|
| $(TMA)_{0.5}Fe_2S_2$ | 4.1 | 13.1 | 1 |

**Table S3.** Fe:Se ratio from EDS normalized to Se.

|  | Fe | Se |
|---|---|---|
| $(TMA)_{0.5}Fe_2Se_2$ | 1.1(1) | 1.0(1) |

The molar ratio of C:H:N was determined to 4.1:13.1:1 which is consistent with $C_4H_{12}N^+$ and confirms the integrity of the TMA$^+$ ion. The CHN elemental analysis was single determined and therfore no standard deviations are given. The NHC mass fraction with respect to FeSe was 12.02 %. This corresponds to a value of 0.24 TMA$^+$ molecules per FeSe and confirms the composition $(TMA)_{0.5}Fe_2Se_2$.

## 5. Magnetic susceptibility and dc resistivity measurements

Magnetic measurements of $\beta$-FeSe and $(TMA)_{0.5}Fe_2Se_2$ were carried out on a Physical Property Measurement System (PPMS-9, Quantum Design) with a vibrating sample magnetometer (VSM). Zero-field cooled and field-cooled measurements were conducted between 2 K and 100 K and an applied field of 15 Oe. The isothermal magnetization was measured at 2 K and 300 K ($H = \pm 50$ kOe). For the resistivity measurements, the samples were ground and pressed into pellets (diameter 5 mm, thickness ~ 0.8 mm). The pellets were contacted with the Wimbush press contact assembly for van der Pauw measurements.[20, 21]

## 6. Infrared spectroscopy

FT-IR spectra were measured on a Bruker Vertex-80V FT-IR spectrometer ($\tilde{\nu}$ = 350 – 4000 cm$^{-1}$). Figure S3 shows the spectra of tetramethylammonium iodide (TMAI) and $(TMA)_{0.5}Fe_2Se_2$ together with the spectrum calculated by DFPT. FeSe is not infrared active. The TMA$^+$ ion has $T_d$ symmetry, and 7 of the 19 fundamental vibrations are infrared active.[22] Raman measurements were not possible due to the strong absorption of the product (black color).

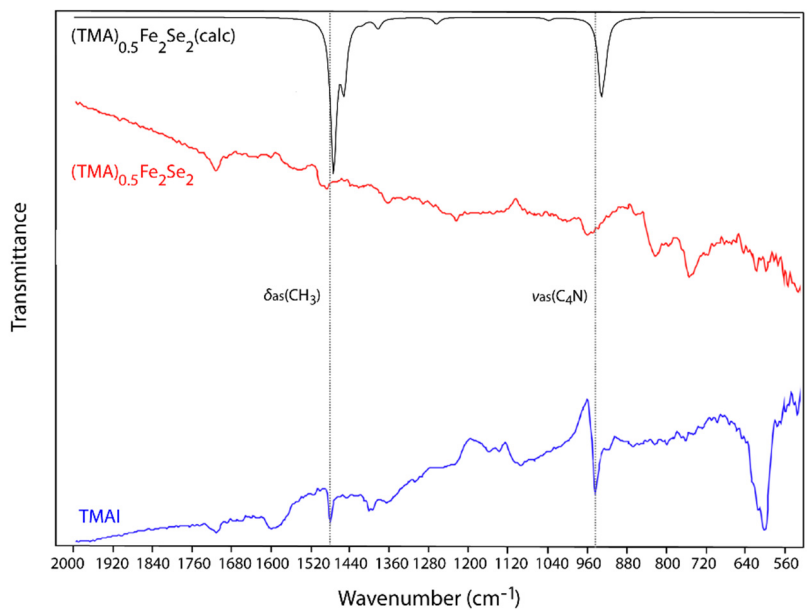

**Figure S3.** FT-IR spectra of TMAI and (TMA)$_{0.5}$Fe$_2$Se$_2$ and the spectrum calculated from DFPT

The TMA$^+$ ions in (TMA)$_{0.5}$Fe$_2$Se$_2$ are located between FeSe layers. Therefore, in comparison to TMAI, the infrared active species in our samples are strongly diluted by the strong IR absorber FeSe. This may be the main factor for the weak intensity of the bands in the spectra. The FT-IR spectra of TMAI and (TMA)$_{0.5}$Fe$_2$Se$_2$ are nevertheless compatible with intercalation of TMA$^+$ into FeSe. This indicates the asymmetric deformation mode vibrations of the methyl group $\delta_{as}$(CH$_3$) at 1481 cm$^{-1}$ and 1501 cm$^{-1}$ in both spectra (TMAI and (TMA)$_{0.5}$Fe$_2$Se$_2$), respectively. The bands are in accordance with literature and only a slight shift to higher wavenumber is apparent (1483 cm$^{-1}$).[23] Furthermore, around 958 cm$^{-1}$ a band is visible in both spectra, which could be assigned to the asymmetric stretching mode of the skeletal C$_4$N. The strong band at ~600 cm$^{-1}$ in the TMAI spectra might be assigned to methyl iodide which could originate from a side reaction during the measurement process. This band is not visible in the product spectrum. Note that the DFPT calculated spectrum matches the measured one well except for a slight zero point shift.

## 7. Deintercalation of (TMA)$_{0.5}$Fe$_2$Se$_2$

A (TMA)$_{0.5}$Fe$_2$Se$_2$ sample was heated to 200 °C for 4 h under argon atmosphere. The residual black powder was analyzed by powder diffraction and magnetic measurements. The powder pattern revealed single phase $\beta$-FeSe (see Figure S4), thus (TMA)$_{0.5}$Fe$_2$Se$_2$ has been quantitatively deintercalated.

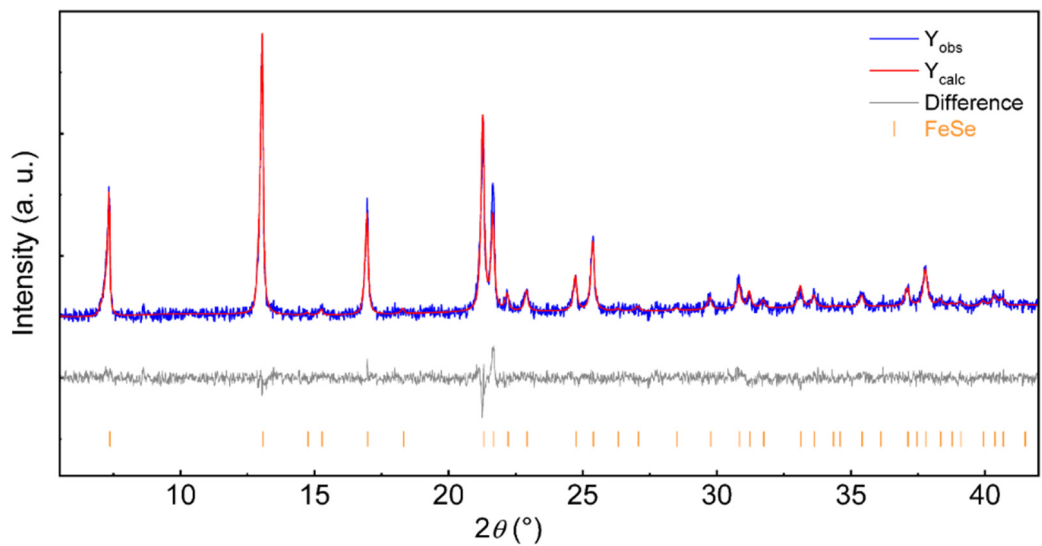

**Figure S4.** PXRD pattern of the residue after heating at 200 °C with Rietveld fit (red) and difference curve (grey).

The regained β-FeSe has the same lattice parameters ($a$ = 3.771(3) Å, $c$ = 5.524(7) Å) as the starting material. The susceptibilities curves show that the regained β-FeSe is superconducting at 8 K, which is consistent with the original properties of β-FeSe (see Figure S5).

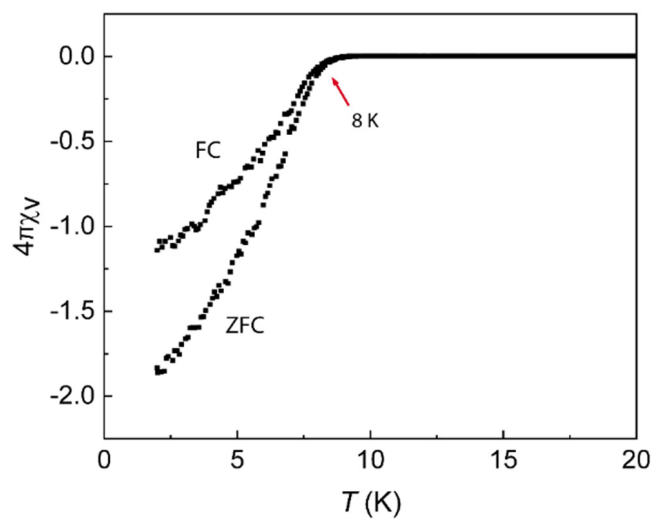

**Figure S5.** Magnetic susceptibility of the residue after heating. FC is field cooled and ZFC is zero-field cooled.

**References (Supporting Information)**

1. Chareev, D.; Osadchii, E.; Kuzmicheva, T.; Lin, J.-Y.; Kuzmichev, S.; Volkova, O.; Vasiliev, A., Single crystal growth and characterization of tetragonal FeSe1−x superconductors. CrystEngComm 2013, 15, (10), 1989–1993.

2. Tambornino, F.; Sappl, J.; Pultar, F.; Cong, T. M.; Hübner, S.; Giftthaler, T.; Hoch, C., Electrocrystallization: A Synthetic Method for Intermetallic Phases with Polar Metal–Metal Bonding. Inorg. Chem. 2016, 55, (21), 11551–11559.

3. Coelho, A., TOPAS-Academic. Version 4.1, Coelho Software: Brisbane, Australia, 2007.

4. Coelho, A. TOPAS Academics, 6; Coelho Software: Brisbane, Australia, 2016.

5. Coelho, A., Indexing of powder diffraction patterns by iterative use of singular value decomposition. J. Appl. Crystallogr. 2003, 36, (1), 86–95.

6. Pawley, G., Unit-cell refinement from powder diffraction scans. J. Appl. Crystallogr. 1981, 14, (6), 357–361.

7. Oszlanyi, G.; Suto, A., The charge flipping algorithm. Acta Crystallogr. Sec. A 2008, 64, (1), 123–134.

8. Brandenburg, K. Diamond, 3.2k; Crystal Impact GbR: Bonn, Germany, 2014.

9. WinXPOW, Version 3.0.2.5; STOE & Cie GmbH: Darmstadt, Germany, 2011.

10. Kresse, G.; Furthmüller, J., Efficient iterative schemes for ab initio total-energy calculations using a plane-wave basis set. Phys. Rev. B 1996, 54, (16), 11169-11186.

11. Kresse, G.; Furthmüller, J., Efficiency of ab-initio total energy calculations for metals and semiconductors using a plane-wave basis set. Comput. Mater. Sci. 1996, 6, (1), 15–50.

12. Kresse, G.; Joubert, D., From ultrasoft pseudopotentials to the projector augmented-wave method. Phys. Rev. B 1999, 59, (3), 1758–1775.

13. Sun, J.; Ruzsinszky, A.; Perdew, J. P., Strongly Constrained and Appropriately Normed Semilocal Density Functional. Phys. Rev. Lett. 2015, 115, (3), 036402.

14. Monkhorst, H. J.; Pack, J. D., Special points for Brillouin-zone integrations. Phys. Rev. B 1976, 13, (12), 5188-5192.

15. Setyawan, W.; Curtarolo, S., High-throughput electronic band structure calculations: Challenges and tools. Comput. Mat. Sci. 2010, 49, (2), 299-312.

16. Stokes, H. T.; Hatch, D. M., FINDSYM: program for identifying the space-group symmetry of a crystal. J. Appl. Crystallogr. 2005, 38, (1), 237-238.

17. Togo, A.; Tanaka, I., First principles phonon calculations in materials science. Scripta Materialia 2015, 108, 1-5.

18. Ganose;, A. M.; Jackson;, A. J.; Scanlon;, D. O., SUMO: Command-line tools for plotting and analysis of periodic ab initio calculations. Journal of Open Source Software 2018, 3, 717.

19. SmartSEM, Version 5.07 Beta; Carl Zeiss Microscopy Ltd.: Cambridge, UK, 2014.

20. QUANTAX 200, Version 1.9.4.3448; Bruker Nano GmbH: Berlin, Germany, 2013.

21. MultiVu, Version 1.5.11; Quantum Design Inc.: San Diego, USA, 2013.

22. Bottger, G. L.; Geddes, A. L., The infrared spectra of the crystalline tetramethylammonium halides. Spectrochim. Acta 1965, 21, (10), 1701–1708.

23. Kornath, A.; Blecher, O.; Neumann, F.; Ludwig, R., Vibrational spectra of the tetramethylpnikogenonium ions. J. Mol. Spectrosc. 2003, 219, (1), 170–174.